\documentclass[10pt]{elsart}

\usepackage{graphicx,amsfonts,amssymb,bm,subeqnarray}

\begin{document}

\begin{frontmatter}

\title{Perfect antireflection via negative refraction}

\author{Juan J. Monz\'on, Alberto G. Barriuso,
Luis L. S\'{a}nchez-Soto}

\address{Departamento de \'Optica,
Facultad de F\'{\i}sica, Universidad Complutense,
28040~Madrid, Spain}

\date{\today}

\journal{Physics Letters A}

\maketitle

\begin{abstract}
We suggest a geometrical framework to discuss
the action of slabs of negatively refracting
materials. We show that these slabs generate
the same orbits as normal materials, but
traced out in opposite directions. This
property allows us to confirm that the
action of any lossless multilayer can be
optically cancelled by putting it together
with the multilayer constructed as the inverted
mirror image, with $\epsilon$ and $\mu$
reversed in sign.
\end{abstract}
\end{frontmatter}

In the last years the notion of materials with
both negative electrical permittivity $\epsilon$
and magnetic permeability $\mu$ is at the center
of a lively and sometimes heated debate. This idea
dates back to 1968, when Veselago~\cite{Ves68}
theoretically predicted that these remarkable
materials would exhibit a number of unusual effects
derived from the fact that in them the vectors
($\mathbf{k}, \mathbf{E}, \mathbf{H}$) of a plane
wave form a left-handed (LH) rather than a right-handed
(RH) set. For this reason, he called them LH media.

The first feasible implementation of such materials
was suggested by Pendry~\cite{Pen96,Pen99}, who
also made the provocative (and criticized)
prediction that they can also act as a perfect
lens~\cite{Pen00}. Inspired by these ideas,
Smith~\textit{et al}~\cite{SPV00} constructed
an artificial medium (consisting of microstructured
arrays of small metallic wires and split ring
resonators) with the desired properties in the
microwave regime. Since then, new samples have
been prepared~\cite{BAO02,LMG03} and several
potential future applications have been
speculated~\cite{Pen03}. To prevent the
significant losses of these metamaterials,
Notomi~\cite{Not00} suggested that identical
behaviors could be expected to occur in
lossless photonic crystals. Many researchers
are now exploring this interesting
possibility~\cite{CAO03,BMS04,MFZ05}.

One of the most interesting properties of these
LH materials is a negative refraction at the
interface with a RH medium. Although this has been
challenged by some authors~\cite{VWV02,PE02}, the
seminal work of Shelby, Smith and Schultz~\cite{SSS01},
as well as other subsequent experiments using
different systems~\cite{PGL03,HBC03}, have dispelled
any doubt regarding the reality of negative refraction.

Most of the work reported in the recent
literature has been focused on the behavior
of the evanescent components. The emphasis on
the near-field limit comes from the interest
that these media evoke as perfect lenses to
transfer images. In this Letter we adopt a
different and simpler strategy and center our
attention in the far-field, much in the standard
transfer-matrix formalism employed when dealing
with multilayers~\cite{Yeh88}. As we have recently
put forward~\cite{YMS02,MYS02}, the action of
the system can be conveniently viewed as a
bilinear transformation on the unit disk.
This geometrical setting allows us to
characterize the slabs by the associated
orbits. It turns out that these orbits are the
same for LH and RH materials, but they are traced
out in opposite directions. This leads to an
intuitive understanding of an intriguing result
obtained by Pendry and Ramakrishna~\cite{PR03}
and rederived recently by Lakhtakia~\cite{Lak02}
and Ruppin~\cite{Rup04}:
a LH slab cancels an identical RH slab.
Furthermore, a much wider class of cancellation
will be confirmed using our approach. We stress
that the transfer matrix solely relies on the
linearity of the wave equation, so that our
treatment applies to any kind of waves and
can seed light into fields where the notion
of bandgap materials is becoming more
and more important, such as sound  or water
waves~\cite{Dow}.

We start by considering the simple example of
a plane parallel slab of thickness $d_1$ and
refractive index $n_1$, surrounded by two
semi-infinite identical media (ambient, $a$, and
substrate, $s$, respectively) of refractive
index $n_0$. For simplicity, all the media are
assumed to be homogeneous, isotropic and lossless.

A monochromatic linearly polarized plane
wave falls from the ambient making an angle
$\theta_0$ with the normal to the first
interface and with an amplitude $E_{a}^{(+)}$.
We consider as well another plane wave of the
same frequency and polarization, and with amplitude
$E_{s}^{(-)}$, incident from the substrate at
the same angle $\theta_0$. The output fields in
the ambient and the substrate will be denoted
$E_{a}^{(-)}$ and $E_{s}^{(+)}$, respectively.

The field amplitudes at each side of this RH
slab are related by the linear relation
\begin{equation}
\label{Evec}
\left [
\begin{array}{c}
E_a^{(+)} \\
E_a^{(-)}
\end{array}
\right ]
= \mathsf{M}_{\mathrm{RH}}
\left [
\begin{array}{c}
E_s^{(+)} \\
E_s^{(-)}
\end{array}
\right ] ,
\end{equation}
where the transfer matrix $\mathsf{M}_{\mathrm{RH}}$
can be explicitly constructed as~\cite{AB87}
\begin{equation}
\label{prod}
\mathsf{M}_{\mathrm{RH}} =
\mathsf{I}_{01}
\mathsf{L}_{1}
\mathsf{I}_{10} .
\end{equation}
Here $\mathsf{I}_{ij}$ accounts for the interface
between the media $i$ and $j$ and has the form
\begin{equation}
\mathsf{I}_{ij} = \frac{1}{t_{ij}}
\left [
\begin{array}{cc}
1 & r_{ij} \\
r_{ij} & 1
\end{array}
\right ] ,
\end{equation}
$t_{ij}$ and $r_{ij}$ being the Fresnel transmission
and reflection coefficients for the interface.

The matrix $\mathsf{L}_{j}$ describes the propagation
through the layer $j$ and is given by
\begin{equation}
\mathsf{L}_{j} =
\left [
\begin{array}{cc}
\exp(i \beta_j) & 0 \\
0 & \exp(- i \beta_j)
\end{array}
\right ] ,
\end{equation}
where $\beta_j = (2 \pi/ \lambda) n_j d_j
\cos \theta_j$ is the slab phase thickness. The
parameter $\lambda$ is the wavelength in vacuo
and $\theta_j$ is the refraction angle in the
layer.

The overall transfer matrix $\mathsf{M}_{\mathrm{RH}}$
results then
\begin{equation}
\mathsf{M}_{\mathrm{RH}} =
\left [
\begin{array}{cc}
1/T_{\mathrm{RH}} &
R^\ast_{\mathrm{RH}}/T^\ast_{\mathrm{RH}} \\
& \\
R_{\mathrm{RH}}/T_{\mathrm{RH}} &
1/T^\ast_{\mathrm{RH}}
\end{array}
\right ] ,
\end{equation}
where $R_{\mathrm{RH}}$ and $T_{\mathrm{RH}}$ are the
reflection and transmission coefficients for the slab:
\begin{eqnarray}
\label{RT}
R_{\mathrm{RH}} & = &
\frac{r_{01} [1 - \exp(- i 2 \beta_1) ]}
{1- r_{01}^2  \exp(- i 2 \beta_1) } , \nonumber \\
& & \\
T_{\mathrm{RH}} & = &
\frac{(1 - r_{01}^2) \exp(- i \beta_1)}
{1- r_{01}^2  \exp(- i 2 \beta_1) } ,
\nonumber
\end{eqnarray}
in such a way that $|R_{\mathrm{RH}}|^2 +
|T_{\mathrm{RH}}|^2 = 1$.

We are often interested in the transformation
properties of field quotients rather than
the fields themselves. Therefore, we
introduce the complex numbers
\begin{equation}
\label{defz}
z  = \frac {E^{(-)}}{E^{(+)}} ,
\end{equation}
for both ambient and substrate. Equation~(\ref{Evec})
defines a transformation on the complex plane
${\mathbb{C}}$, mapping the point $z_s$ into
the point $z_a$, according to
\begin{equation}
\label{accion}
z_a =  \frac{\mathfrak{b}^\ast + \mathfrak{a}^\ast z_s}
{\mathfrak{a}  + \mathfrak{b} \, z_s} ,
\end{equation}
where $\mathfrak{a} = 1/T_{\mathrm{RH}}$ and
$\mathfrak{b} = (R_{\mathrm{RH}}/T_{\mathrm{RH}})^\ast$.
The matrix element $\mathfrak{b}$ is always a
imaginary number for a symmetric system (i.e., a system
for which the reflection and transmission coefficients
are the same whether light is incident on  one side or
on the opposite side). Equation~(\ref{accion}) is a
bilinear (or M\"{o}bius) transformation and one can
check that the unit disk remains invariant under the
slab action~\cite{YMS02}. Henceforth we assume that no
light strikes from the substrate and then we have $z_s = 0$
and $z_a = R_{\mathrm{RH}}$.

\begin{figure}
\centering
\includegraphics[height=5cm]{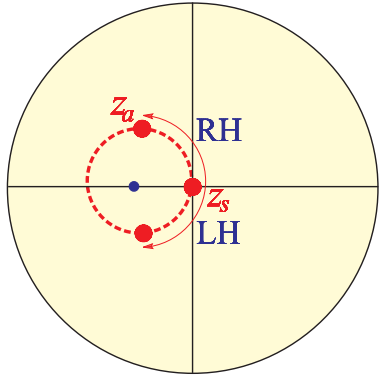}
\caption{Plot of a typical orbit in the unit disk for
a slab. The counterclockwise path is associated with
the RH slab, while the clockwise one corresponds to the
LH slab. To compute the transformed point $z_a$, we have
taken $n_1 = 1.75$, $n_0=1$, $\lambda=1 \ \mu$m and $\theta_0=
45^\circ$. We have also marked the fixed point.}
\end{figure}

For this slab $ [\mathrm{Tr} (
\mathsf{M}_{\mathrm{RH}} )] ^2 < 4$ and the
action in the unit disk leaves only one point
invariant (fixed point)~\cite{SMY01}.  To picture
how $z_s$ transforms into $z_a$ the concept of orbit
is especially appropriate. Given the point $z_s$,
its orbit is the set of points $z^\prime$ obtained
from $z_s$ by the action of the family of matrices
representing a slab. This family can be generated,
e.g., by varying continuously the thickness. One
can then show that the orbits obtained are always
(hyperbolic) circumferences centered at the fixed
point and passing through the point $z_s = 0$.
In Fig.~1 we have plotted a typical orbit as well
as the transformed of the point $z_s = 0$ for a
slab of phase thickness of $\beta_1 = 0.8750 \pi$ rad.

Let us consider the same slab, but of a LH medium,
with negative refractive index $- n_1$. The same
arguments used to assume a negative index lead to
the conclusion that a positive wave impedance;
i. e., $Z = \sqrt{\mu/\epsilon}$, is the correct
choice. This translates into the fact that the
Fresnel equations remain valid provided the
absolute values of $\epsilon$ and $\mu$ are used.
On the other hand, it is well confirmed that the
phase velocity is oppositely directed to the
energy flow in these media.

All this together means that the interface matrices
$\mathsf{I}_{ij}$ are the same as for the corresponding
RH slab, while the layer matrices become complex
conjugate. In other words, the matrix
$\mathsf{M}_{\mathrm{LH}}$ for this slab is
\begin{equation}
\label{M010}
\mathsf{M}_{\mathrm{LH}} =
\mathsf{I}_{01}
\mathsf{L}^\ast_{1}
\mathsf{I}_{10} =
\mathsf{M}^\ast_{\mathrm{RH}} =
\mathsf{M}^{-1}_{\mathrm{RH}} .
\end{equation}
From this apparently innocuous formula, one can
draw several nontrivial and interesting conclusions.
First, we note that if we plug the matrix elements
of $\mathsf{M}_{\mathrm{LH}}$ in the bilinear
action~(\ref{accion}), the fixed point is the same,
but the orbit is traced out in opposite direction.
Therefore, in this geometrical picture, LH and RH materials
have identical orbits, although for the former
they are clockwise, while for the latter they are
counterclockwise.

Let us now put together these RH and LH slabs. The
resulting system is described by the product of the
transfer matrices $\mathsf{M}_{\mathrm{RH}}$ and
$\mathsf{M}_{\mathrm{LH}}$, which, by virtue of
Eq.~(\ref{M010}), is precisely the identity.
In consequence, we get a perfect antireflector
with no phase change in transmission. This is
quite intuitive from our unit-disk picture:
the action of the global system consists of two
successive identical rotations in opposite directions
that cancel out.

\begin{figure}
\centering
\includegraphics[height=4cm]{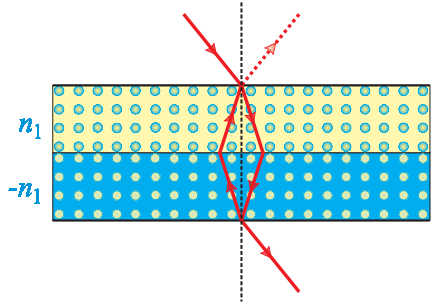}
\caption{Scheme of the energy flow for the system
resulting by putting together two identical slabs, one made
of RH and the other of LH material. Both constitute a pair
of complementary media, each cancelling the effect of the other.}
\end{figure}

Alternatively, one can look at this problem by
using the usual and intuitive method of adding
multiply reflected and transmitted waves.
Since in the interface between both slabs there
is no reflected wave, the scheme of the energy flow
is as indicated in Fig.~2. If we take the incident
field of unit amplitude, the overall reflected field is
\begin{equation}
r_{01} + t_{01} t_{10} ( r_{10} +
r_{10}^3 + r_{10}^5 + \ldots ) = 0 ,
\end{equation}
while the overall transmitted field is
\begin{equation}
t_{01} t_{10} (1 + r_{10}^2 +
r_{10}^4 +  \ldots ) = 1 ,
\end{equation}
which confirms the previous result.

The discussion so far admits a straightforward
generalization for any multilayer. Indeed, let
$\mathsf{M}_{as}$ denote the transfer matrix of
a system consisting of an arbitrary number
of layers (some of them made of RH materials and
some of LH materials), which can be constructed
by a direct extension of (\ref{prod}). One can
show that
\begin{equation}
\mathsf{M}_{as} =
\left [
\begin{array}{cc}
1/T_{as} & R^\ast_{as}/T^\ast_{as} \\
& \\
R_{as}/T_{as} & 1/T^\ast_{as}
\end{array}
\right ] ,
\end{equation}
with $|R_{as} |^2 + | T_{as} |^2 = 1$.
Now we take the multilayer in the reverse
order, which is represented by
\begin{equation}
\mathsf{M}_{sa} =
\left [
\begin{array}{cc}
1/T_{as} & - R_{as}/T_{as} \\
& \\
- R^\ast_{as}/T^\ast_{as} & 1/T^\ast_{as}
\end{array}
\right ] .
\end{equation}
Next we switch every RH layer to an identical LH
layer and viceversa. The final system is thus
described by $\mathsf{M}^\ast_{sa}$, and one can
check that~\cite{MYS02}
\begin{equation}
\mathsf{M}^\ast_{sa} = \mathsf{M}^{-1}_{as} .
\end{equation}
In consequence, when both multilayers are put together
they give the identity. This formalizes in a
different framework the notion of ``complementary media"
introduced by Pendry and Ramakrishna~\cite{PR03}:
any medium can be optically cancelled by an equal
thickness of material constructed to be an inverted
mirror image of the medium, with $\epsilon$ and $\mu$
reversed in sign. That is, complementary media cancel
one another and become invisible (i.e., a perfect
antireflector).

In summary, we have demonstrated another curious
property of LH materials, which may be experimentally
tested with the state of the art in this hot area of
research. Although these results could have practical
consequences, in our view they provide the first
feasible implementation of how to build the inverse
of a transfer matrix.

\end{document}